%% file: Game.Process.Survey.GAS2016.tex
\documentclass{sig-alternate}
\usepackage{balance}
\usepackage[utf8]{inputenc}
\usepackage[T1]{fontenc}

\usepackage{tabularx}
\usepackage{rotating}
\usepackage{subfigure}
\usepackage{rotating}
\usepackage{booktabs} 
\usepackage{makecell}

\usepackage{color}

\usepackage[numbers,sort&compress]{natbib} 

\begin{document}

\doi{xxx}
\isbn{978-1-4503-4160-8}
\conferenceinfo{GAS '16}{May 16, 2016, Austin, TX, USA}
\acmPrice{\$zzz}

\title{Are the Old Days Gone? A Survey on Actual Software Engineering Processes in Video Game Industry}

\numberofauthors{2}

\author{
\alignauthor
Cristiano Politowski, Lisandra Fontoura\\
       \affaddr{Federal University of Santa Maria}\\
       \affaddr{Santa Maria, Brazil}\\
       \email{\{cpolitowski,lisandra\}@inf.ufsm.br}
\alignauthor
Fabio Petrillo, Yann-Gaël Guéhéneuc\\
       \affaddr{École Polytechnique de Montréal}\\
       \affaddr{Montréal, Canada}\\
       \email{fabio@petrillo.com, yann-gael.gueheneuc@polymtl.ca}
}
\date{22 January 2016}

\hyphenation{post-mor-tems}
\maketitle

\input{abstract}


\input{introduction}
\input{background}
\input{methodology}
\input{results}
\input{relatedwork}
\input{acknowledgment}
\input{conclusions}

\bibliographystyle{abbrv}
\bibliography{Game.Process.Survey.GAS2016}
\balance

\end{document}

%% file: abstract.tex
\begin{abstract}

In the past 10 years, several researches studied video game development process who proposed approaches to improve the way how games are developed. These approaches usually adopt agile methodologies because of claims that traditional practices and the waterfall process are gone. However, are the ``old days" really gone in the game industry? 

In this paper, we present a survey of software engineering processes in video game industry from postmortem project analyses. We analyzed 20 postmortems from Gamasutra Portal. We extracted their processes and modelled them through using the Business Process Model and Notation (BPMN).

This work presents three main contributions. First, a postmortem analysis methodology to identify and extract project processes. Second, the study main result: \textbf{the ``old days" are gone, but not completely}. \textbf{Iterative practices} are increasing and are applied to at least \textbf{65\% of projects} in which \textbf{45\% of this projects} explicitly adopted Agile practices. However, \textbf{waterfall} process is still applied at least \textbf{30\% of projects}. Finally, we discuss some implications, directions and opportunities for video game development community.

\end{abstract}

%% file: introduction.tex
\section{Introduction}
\label{sec:introduction}

Game development is an extremely complex activity \cite{game_plan}. A study on game industry problems \cite{Petrillo2009} concluded that game projects does not suffer from technological problems but essentially from management and process. To alleviate theses issues, several academic and partitioner studies have been conducted for years, especially on the adoption of agile game development \cite{Kanode2009,Keith2010,Petrillo2010}. Moreover, a recent systematic literature review \cite{OsborneOHagan2014} observed that there are substantially fewer industrial studies about game development processes and claims that agile processes are appropriate when innovation and speed to market are vital in game development. In the same direction, a developer gave us some interesting observations in a recent postmortem about game development \cite{Fridley2013}:

\begin{quotation}
\noindent
\textit{``\textbf{The old days are gone}. You can't expect producers or leads to come up with a huge waterfall of everything they thought would get done over the next three years. In the game development business, it's insane to think you have any insight into what your team will be doing one year from now. You can set major milestones with hard dates, but filling in all the details between those points is an exercise in futility.''}
\end{quotation}

Nevertheless, is this claim general or the developer's bias vision? \textbf{Are ``the old days" really gone in video game industry?}  
For the purpose of answering this question and studying game development processes in industrial context, we conduct a survey of the software engineering processes in video game industry from postmortem project analyses, modelling them using Business Process Model and Notation (BPMN) \cite{Omg2011}. 

This paper is structured as follows. Section \ref{sec:background} summarizes briefly the main concepts and definitions about game development processes and practices used in this work. Section \ref{sec:methodology} presents our research methodology to analyse and collect data from project postmortems. Section \ref{sec:results} shows the results of study, providing project process diagrams, quantitative and qualitative results, and discussing our findings. Section \ref{sec:relatedwork} presents the related work. Section \ref{sec:threats} shows threats of validity. Finally, Section \ref{sec:conclusions} summarizes this survey and discusses some implications and future work opportunities.

%% file: background.tex
\section{Background}
\label{sec:background}

In this section, we define briefly some concepts used in this work to provide a single interpretation. First, we define concepts about game development processes described in the literature. Then we describe postmortems in game development.

\subsection{Game Development Process}
\label{gamedev_process}

Game development processes can be classified into four main categories: waterfall, iterative, hybrid, and ad-hoc \cite{Al-azawi2014, OsborneOHagan2014}. In this work, we define each category as follows:

\textbf{Waterfall} \cite{royce1970managing} (or predictive) is a sequential process in which a next phase is started only if the previous phase is completely finished, delivering business value all at once. This is the traditional game development process, requiring explicit requirement assessments followed by orderly and precise problem solving procedures.

\textbf{Iterative} \cite{larman2003iterative} is a process which consists to develop a software by repeating short-cycles to  deliver a ready-to-use feature each time. Agile software methodology follows this iterative approach, improving  continuously and systematically its processes and practices \cite{fowler2001new}.

\textbf{Hybrid} is a combination of waterfall and iterative processes in the same project. Typically, the waterfall strategy is used during pre/post-production and the iterative is applied during the production phase.

\textbf{Ad-Hoc} is a process that is created only for a specific project, without a previous definition. In ad-hoc process, activities are defined on demand and the process changes to respond to punctual and contextual issues.

\subsection{Game Development Postmortems}
\label{sec:postmortems}

The term \textit{postmortem} designates a document that summarizes the project development experiences, with a strong emphasis on the positive and negative outcomes of the development \cite{goodbye_postmortem}. It is commonly done right after the project finishes, by managers or senior project participants \cite{callele}. It is a important tools for knowledge management \cite{postmortem}, from which the group can learn from its own experiences and plan future projects. Postmortem analysis can be so revealing that some authors \cite{postmortem} argue that no project should finish without postmortem.

Postmortems are much used in the game industry. Many game Web sites devote entire sections to present these documents, such as \textit{Gamasutra} (http://www.gamasutra.com) and \textit{Gamedev} (http://www.gamedev.net). These postmortems pertain to a variety of development teams profiles and projects, varying from few developers in small projects to dozens of developers in five-year-long projects.

The postmortems published by {\em Gamasutra} mainly follow the structure proposed by the \textit{Open Letter Template} \cite{pma}, which is composed of three sections. The first section summarizes the project and presents some important outcomes of the development. The next two sections discuss the most interesting of game development:

\begin{itemize}
	
	\item \textbf{What went right:} it discusses the {\em best practices} adopted by developers, solutions, improvements, and project management decisions that have improved the efficiency of the team. All these aspects are critical elements to be used in planning future projects.
	
	\item \textbf{What went wrong:} it discusses difficulties, pitfalls, and mistakes experienced by the development team in the project, both technical and managerial.
	
\end{itemize}

The information contained in postmortems constitute a piece of knowledge that can be reused by any development team, including examples and real life development experiences. They allow knowledge sharing and are useful for planning future projects or process improvements. 


%% file: methodology.tex
\section{Research Methodology}
\label{sec:methodology}

\begin{figure*}
	\centering
	\includegraphics[width=1\linewidth]{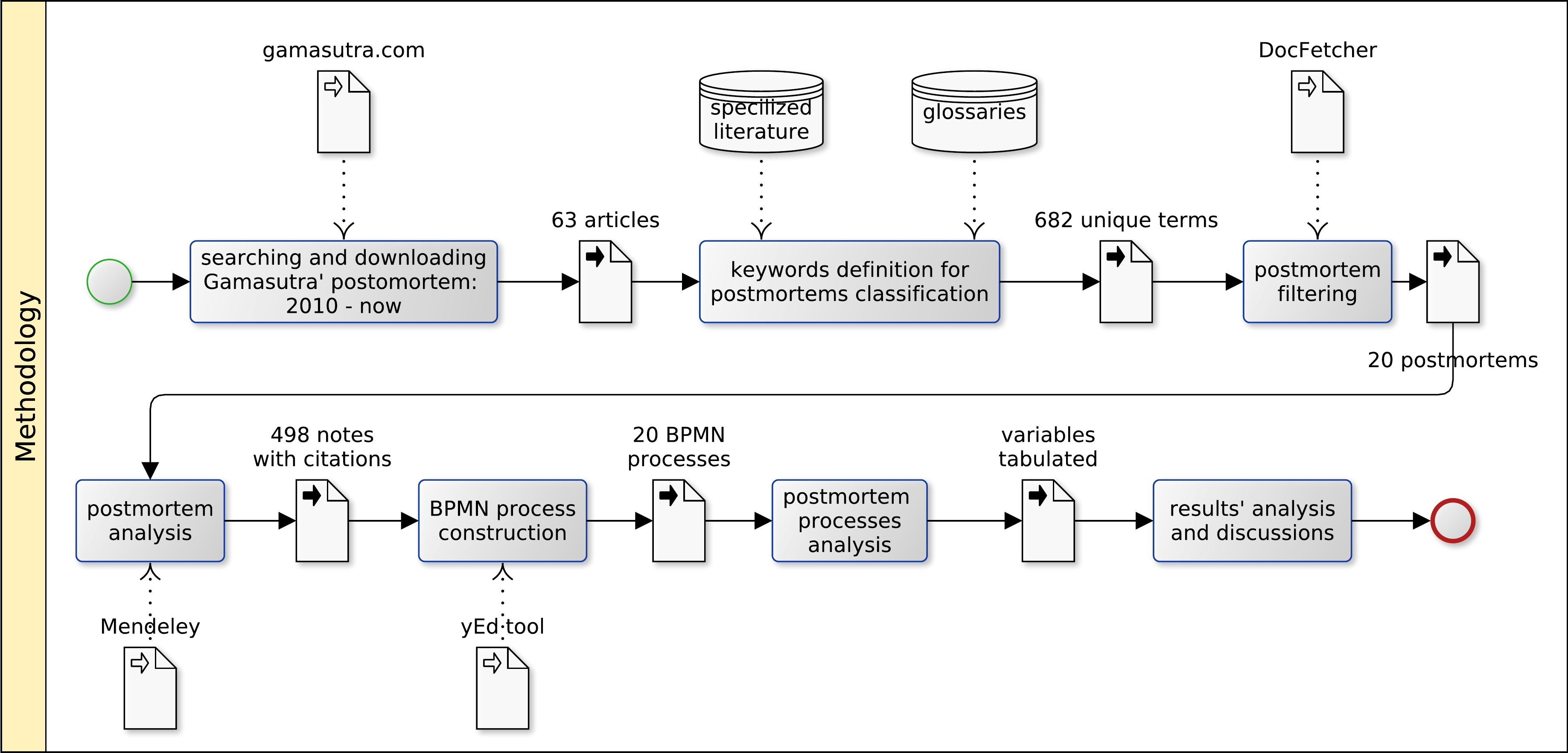}
	\caption{The research method in BPMN notation.}
	\label{fig:methodology}
\end{figure*}

\begin{figure*}[t]
	\centering
	\includegraphics[width=1\linewidth]{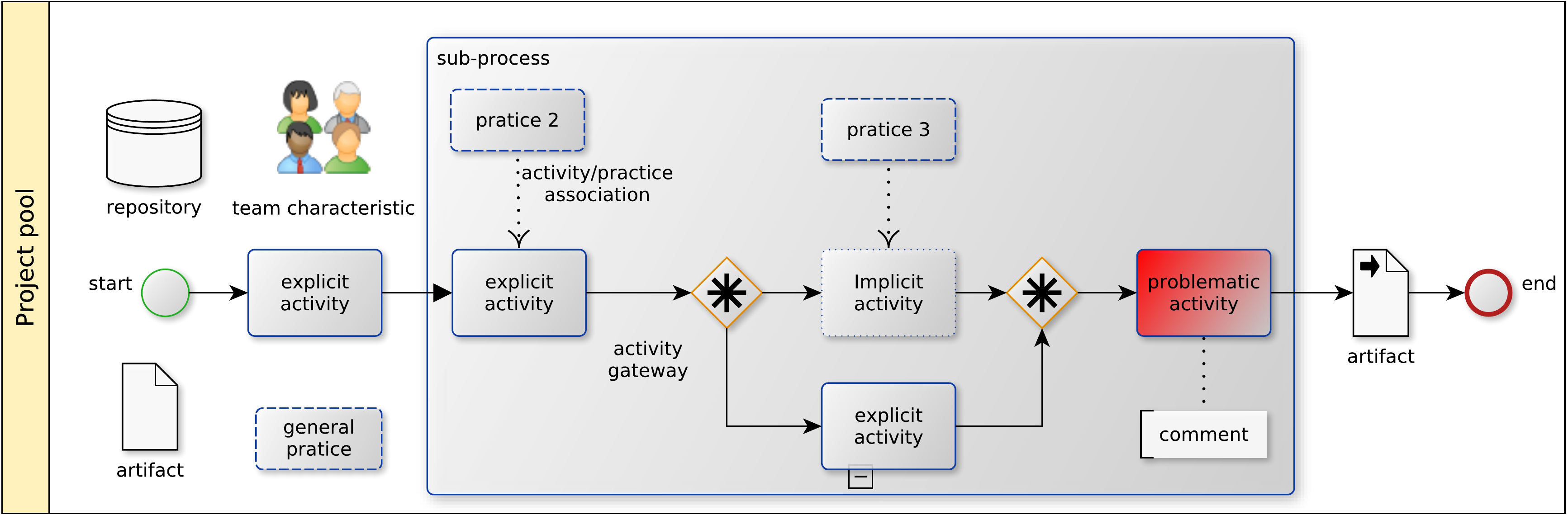}
	\caption{BPMN meta-model.}
	\label{fig:metamodel}
\end{figure*}

\begin{table}[h]
	\scriptsize
	\centering
	\caption{List of analyzed postmortems and notes.} 
	\begin{tabularx}{\linewidth}{@{}c X c@{}}
		\hline
		\textbf{\#} & \textbf{Postmortems}                       & \textbf{\# notes} \\ \hline
		1           & Brutal Legend                              & 16                \\
		2           & Kingdoms of Amalur: Reckoning              & 19                \\
		3           & Caseys Contraptions                        & 18                \\
		4           & Sins of a Solar Empire                     & 27                \\
		5           & Amnesia: A Machine for Pigs                & 27                \\
		6           & City Conquest                              & 26                \\
		7           & Baldurs Gate Enhanced Edition              & 26                \\
		8           & Trine                                      & 49                \\
		9           & Natural Selection 2                        & 28                \\
		10          & The Path                                   & 35                \\
		11          & Dust An Elysian Tail                       & 28                \\
		12          & Anomaly Warzone Earth                      & 42                \\
		13          & Aaaaa! -- A Reckless Disregard for Gravity & 14                \\
		14          & Scooby-Doo First Frights                   & 25                \\
		15          & Spider-Man                                 & 32                \\
		16          & Deadliest Warrior                          & 12                \\
		17          & Zack Zero                                  & 23                \\
		18          & God of War Ascension                       & 22                \\
		19          & Electronic Symphony The Untold Story       & 8                 \\
		20          & Guacamelee                                 & 21                \\ \hline
		& Total                                      			 & 498               \\ \hline
	\end{tabularx}
	\label{tab:pm_notes}
\end{table}

To conduct our survey and answer our research question, we elaborated a research method to analyze postmortems and extract the processes used in game development. Our method contains several steps whose output is the input for a next step. In this section, we present these steps, which are modeled in Figure \ref{fig:methodology}.

\textbf{Step 1. Getting postmortems:} 
first, we searched in Gamasutra by all the postmortems ranged from 2010 until now. Our focus was gather the latest projects, so postmortems prior to 2010 were discarded. We collected a total of 63 articles\footnote{The collection of our raw data and diagrams are available in http://gas2016.github.io}. 

\textbf{Step 2. Keyword definition:} as done by Petrillo \textit{et al.} \cite{Petrillo2009}, we limited  our study in 20 postmortems, defining keywords for filtering, searching by concepts on software engineering \cite{martin2003agile, martin2009clean, beck2000extreme, cohn2010succeeding, shore2007art, fowler2001new, swebok2014, Kruchten:2000:RUP} and glossaries \cite{ieeeglossary1990,glossary1,glossary2,glossary3,glossary4,glossary5,glossary6}. After that, we got 682 unique terms. 

\textbf{Step 3. Postmortems filtering:} 
next, utilizing the text mining software DocFetcher\footnote{http://docfetcher.sourceforge.net/en/index.html} on the terms got from step 2, we selected the 20 postmortems (Table \ref{tab:pm_notes}) with greater number of found terms.

\textbf{Step 4. Postmortem analysis:} 
with a collection of selected postmortems, we read every postmortem thoroughly, searching for process elements like \textit{activities}, \textit{roles}, \textit{artifacts}, \textit{practices} and all other elements related to software engineering process and management. As a result, we extracted a total of 498 notes and respective quotations from the 20 postmortem analyses. 

\textbf{Step 5. Game process modeling:}
we analyzed the postmortem notes to build a process model for each project, modeling activities or process details, using the Business Process Model Notation (BPMN) version 2.0 \cite{Omg2011}. To guide this modeling process, we defined a meta-model in Figure \ref{fig:metamodel}. Following this steps, we created a process model for each postmortem in which we can observe clearly its characteristics and highlights the work-flow for each project. The meta-model consist in several elements, described on Table~\ref{tab:elements}. Furthermore, we highlighted in red some \textbf{problematic elements} explicitly declared by a postmortem author. 

\textbf{Step 6. Postmortem Process Analysis:} using the process diagrams and quotations, we analyzed each project, establishing two essential variables to answer our research question. The first variable is \textbf{process}, which shows in what process category (Section \ref{gamedev_process}) a process is classified. The second is \textbf{agile}, a boolean variable that shows if a project follows agile practices. Finally, we collected process data and organized them in Table \ref{tab:pm_details}.

\begin{table}
	\scriptsize
	\centering
	\caption{Description of meta-model elements.} 
	\begin{tabularx}{\linewidth}{@{}p{0.17\linewidth} X p{0.20\linewidth}@{}}
		\hline
		\textbf{Element} & \textbf{Description} & \textbf{Examples} \\
		\hline
		\textbf{Team characteristics} & Team characteristics concern team peculiarities. Any detail, highlighted by the author, about the team in development process. & \textit{multi-disciplinary teams}, \textit{novice team} and \textit{full-stack developer}. 
		\\ \Xhline{0.1pt}
		\textbf{Sub-process} & Sub-process highlights a division or a special feature in a workflow. For example, an iterative sub-process represents that the flow was repeated many times. When the word ``production'' is written on a element label, it means a waterfall-like method. & \textit{pre-production}, \textit{production (iterative)}, or others ad-hoc workflows.
		\\ \Xhline{0.1pt} 
		\textbf{Repositories}         & Repositories are used whenever a source of information or a set of ideas was stored. & \textit{idea pool} and \textit{bug repository}. 
		\\ \Xhline{0.1pt}
		\textbf{Artifacts}            & Artifacts are pieces of work that were created, modified, or used during an activity, and it defines an area of responsibility \cite{kruchten2004rational}. Normally, it is an output of an activity or a document used during a process. & \textit{game design document}, \textit{incremental build}, or \textit{assets}. 
		\\ \Xhline{0.1pt} 
		\textbf{Activities}          & Activities, tasks or steps describe units of work that provide a result. They can be  \textit{explicit} (explicitly declared) or \textit{implicit} (inferred by the authors). & \textit{programming}, \textit{prototyping}, and \textit{design}. 
		\\ \Xhline{0.1pt}
		\textbf{Practices}            & Practices are patterns or systematic habits used by teams during a development cycle. They can be associated with an activity. & \textit{TDD}, \textit{collaborative development} and \textit{cutting features}.
		\\ \Xhline{0.1pt}
		\textbf{Activity gateways}    & Activity gateways specify moments from start and end of parallel activities. & 
		\\ \Xhline{0.1pt} 
		\textbf{Association}          & Association arrow are used when some elements are associated to one. For example, to make a comment about something in the process.                                                                                                                         &  
		\\ \Xhline{0.1pt} 
		\textbf{Comments}             & Comments are additional information about some element or event. & 
		\\ \hline
	\end{tabularx}
	\label{tab:elements}
\end{table}

Following these steps systematically, we produced several processes used in the video game industry. These results and their discussions are presented in the next section.

%% file: results.tex
\section{Results}
\label{sec:results}

\begin{figure}[h]
	\centering
	\includegraphics[width=1\linewidth]{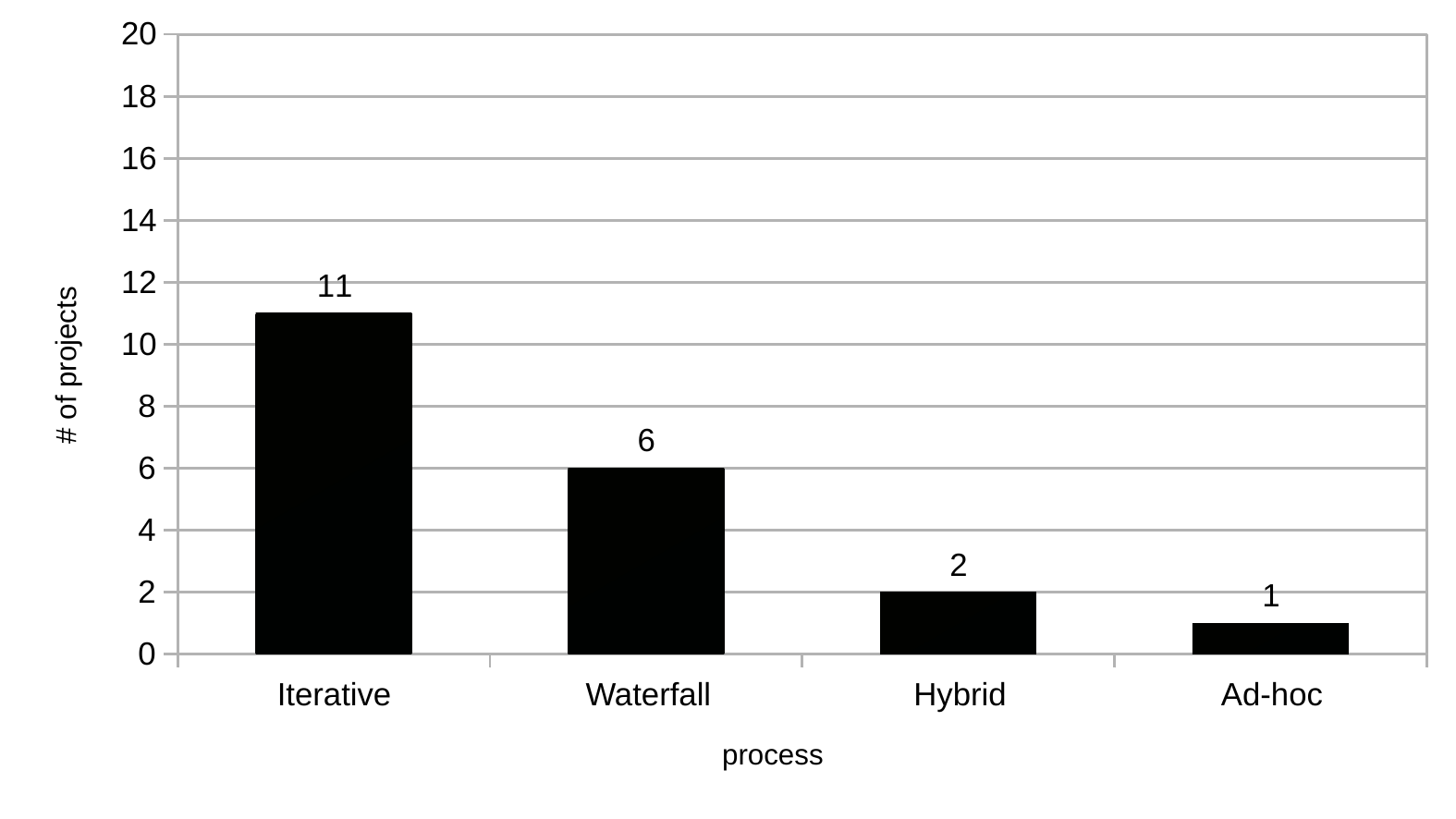}
	\caption{Process occurrences by category}
	\label{fig:chart_process}
\end{figure}

We analyze the postmortem diagrams and quotations, categorizing each project as \textit{waterfall, iterative, hybrid, or ad-hoc} based on the definitions presented in Section \ref{sec:background}. We organize this information in Table \ref{tab:pm_details}. Evaluating this data, we found that 55\% of projects (11/20) adopt an iterative process; 30\% of projects (6/20) use waterfall process; 10\% of projects (2/20) are hybrid; 5\% (1/20) apply ad-hoc practices. These results are presented in Figure \ref{fig:chart_process}. 


\begin{table}
	\scriptsize
	\centering
	\caption{Analyzed postmortem}
	\label{tab:pm_details}
	\begin{tabularx}{\linewidth}{@{}X c c@{}}
		\hline
		\textbf{Postmortem}                  & \textbf{Process} & \textbf{Agile} \\ \hline
		Brutal Legend                        & hybrid           & yes            \\
		Kingdoms of Amalur: Reckoning        & iterative        & yes            \\
		Caseys Contraptions                  & iterative        & yes            \\
		Sins of a Solar Empire               & iterative        & yes            \\
		Amnesia: A Machine for Pigs          & iterative        & yes            \\
		City Conquest                        & iterative        & yes            \\
		Baldurs Gate Enhanced Edition        & iterative        & yes            \\
		Trine                                & waterfall        & no             \\
		Natural Selection 2                  & iterative        & yes            \\
		The Path                             & iterative        & no             \\
		Dust An Elysian Tail                 & waterfall        & no             \\
		Anomaly Warzone Earth                & iterative        & no             \\
		A Reckless Disregard for Gravity     & ad-hoc           & no             \\
		Scooby-Doo First Frights             & waterfall        & no             \\
		Spider-Man                           & hybrid           & yes            \\
		Deadliest Warrior                    & waterfall        & no             \\
		Zack Zero                            & waterfall        & no             \\
		God of War Ascension                 & iterative        & no             \\
		Electronic Symphony                  & waterfall        & no             \\
		Guacamelee                           & iterative        & no             \\ \hline
		
	\end{tabularx}%
\end{table}

\subsection{Typical Process Models}

\begin{figure*}
	\centering
	\includegraphics[width=1\linewidth]{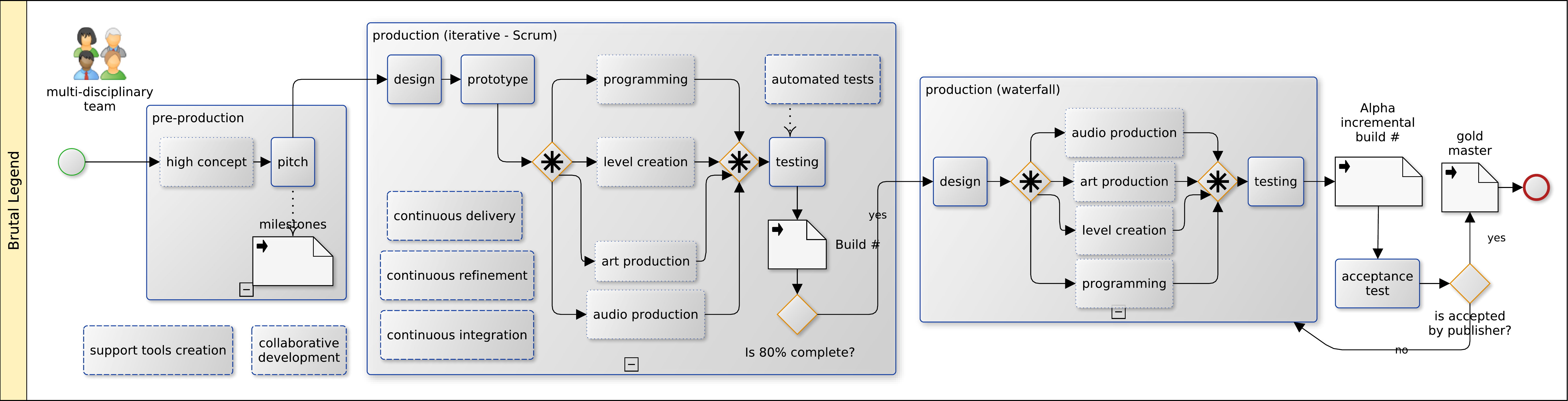}
	\caption{Hydrid process - Brutal Legend model}
	\label{fig:process-brutal_legend}
\end{figure*}

\begin{figure*}
	\centering
	\includegraphics[width=1\linewidth]{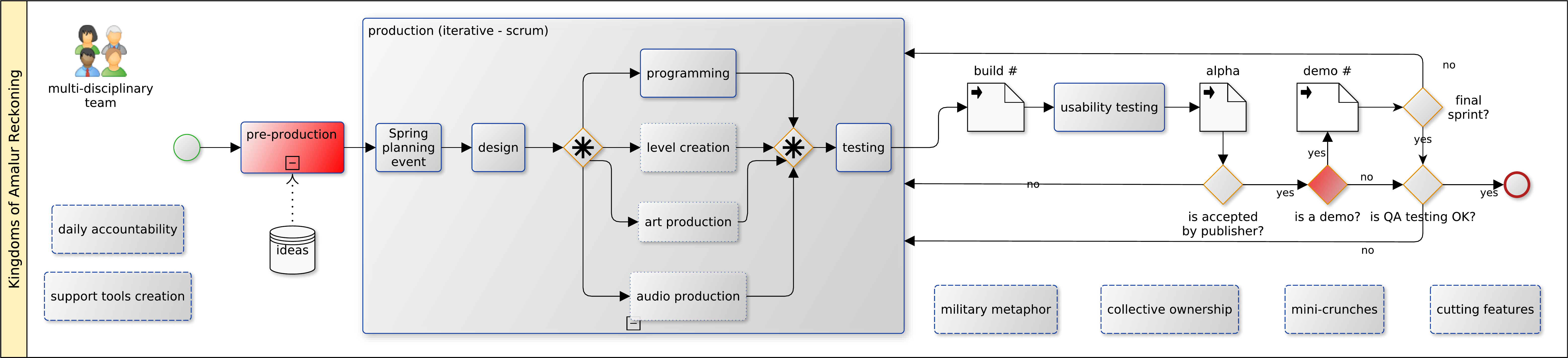}
	\caption{Iterative process - Kingdoms of Amalur: Reckoning model}
	\label{fig:process-kingdoms_of_amalur}
\end{figure*}

\begin{figure*}
	\centering
	\includegraphics[width=1\linewidth]{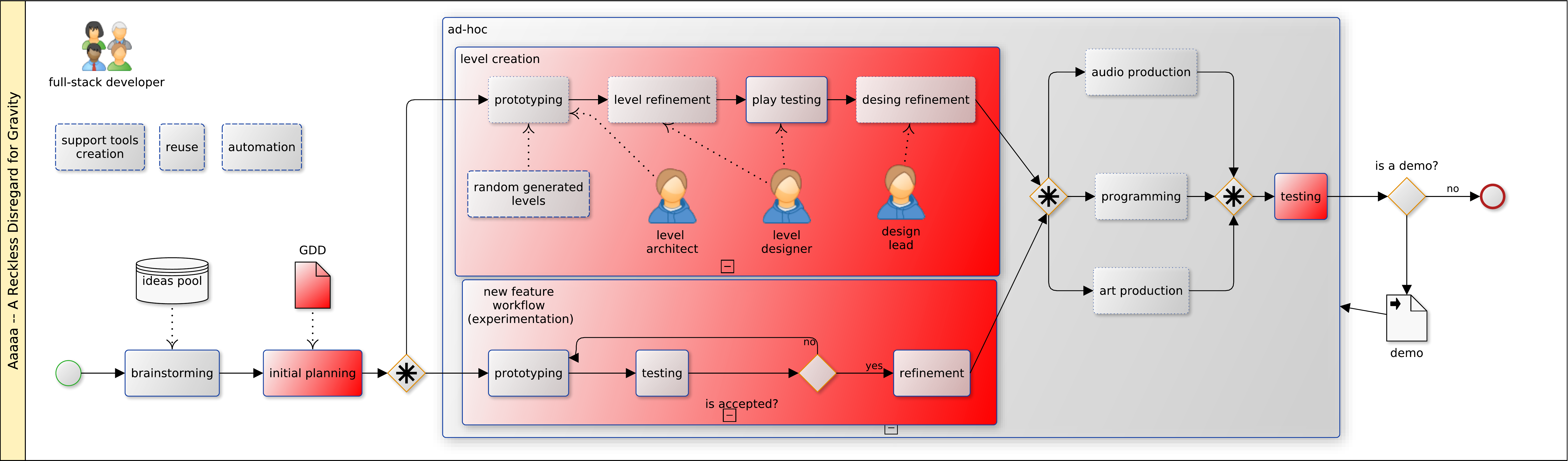}
	\caption{Ad-hoc process: Aaaa! A Reckless Disregard for Gravity model}
	\label{fig:process-a_reckless_disregard_for_gravity}
\end{figure*}

\begin{figure*}
	\centering
	\includegraphics[width=1\linewidth]{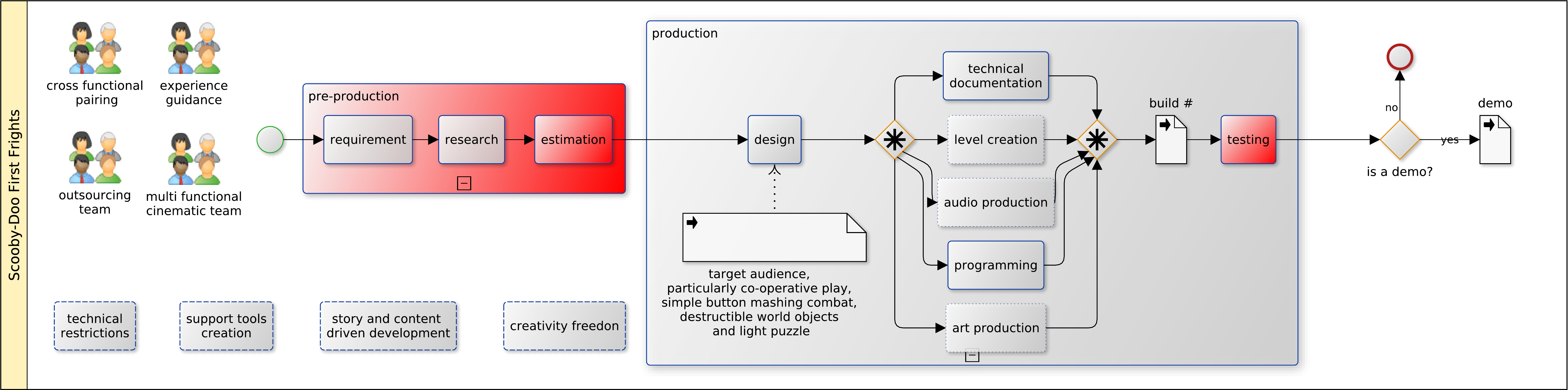}
	\caption{Waterfall process: Scooby-Doo model.}
	\label{fig:process-scooby_doo}
\end{figure*}

To present each process category, we selected four typical project models: a hybrid (Brutal Legend - Figure \ref{fig:process-brutal_legend}), an iterative (Kingdoms of Amalur: Reckoning - Figure \ref{fig:process-kingdoms_of_amalur}) , a ad-hoc (Aaaaa! A Reckless Disregard for Gravity - Figure \ref{fig:process-a_reckless_disregard_for_gravity}), and finally a waterfall (Scooby-Doo First Frights - Figure \ref{fig:process-scooby_doo})\footnote{All 20 postmortem process models are available on http://gas2016.github.io}. The next paragraphs highlight some common characteristics typically found in each process category.

Composed by a combination of iterative and waterfall processes, Figure \ref{fig:process-brutal_legend} is an example of hybrid process. This process has a remarkable characteristic: the process is clearly separated in pre-production, production, and post-production. The pre-production is composed by high concepts of the game, pitch construction, and milestones definition. Once this phase finished, development comes to the iterative production phase, where a set of activities are performed repeatedly, delivering a playable and testable game build in every iteration. When the project is nearly to be done,  the team change for the waterfall approach, producing the remaining features and integrating all parts to deliver a gold version. Some agile practices are typically used like collaborative development, automated testing and continuous delivery. 

Iterative process is defined by development cycles. \textit{Kingdoms of Amalur: Reckoning} is a typical example of this process, modeled in Figure \ref{fig:process-kingdoms_of_amalur}. The pre-production phase is usually short, producing a game main vision and decisions, followed by several complete cycles of development. During iterations, agile methodologies and practices (like Scrum \cite{Keith2010}) are applied by 45\% of analyzed projects. Some significant activities are sprint planning events, continuous delivering, and testing. 

Ad-hoc process is exemplified in Figure \ref{fig:process-a_reckless_disregard_for_gravity}. In this case, the development start with brainstorming and initial planning activities with use of game design document. The production can be separated by level creation, experimentation and testing. Tasks are distributed by role, following a contextual sequence. This approach is usually adopted by small teams.

Waterfall process consists of well-separated sequence of phases, showing in Figure \ref{fig:process-scooby_doo}. First, a game conception is made during the pre-production phase. Second, a complete game design is create during the design phase. Third, using this design, a game is implemented (programming, audio, art). There are not intermediate deliverables or scheduled milestones. Next, the game made is tested. Finally, the complete product is delivered. Waterfall process has usually well-defined multi-functional teams.

\subsection{Result Analyzes}

Analyzing these models and results, we conclude that \textbf{the ``old days" are gone, but not completely}. Despite some particularities discussed by Murphy-Hill \textit{et al.} \cite{Murphy-Hill2014}, \textbf{this work shows that actually video game and traditional software development share similar processes and practices}. The video game industry has improved its processes, adopting regular software engineering techniques. In the same direction of previous academic studies, \textbf{we found that iterative process is actually \textit{mainstream} in video game industry and agile practices adoption is increasing in the last years}. However, we believe that iterative process and agile practice benefits are yet \textbf{misunderstood} by some game developers, managers, producers, publishers, and educators. 

Furthermore, significant facts were found on \textit{publishers} and \textit{milestones}. Once a company (including indie) made a deal with a publisher, normally a set of milestones are predefined, and when a portion of the game must be presented. These milestones are usually underestimated,  It can, most of the time, usually . This situation was observed in \textit{Anomaly Warzone Earth} postmortem:

\begin{quotation}
\noindent
\textit{``Knowing from the past that chasing milestones could leave us with heavy crunch time, which was something that no one wanted at all, we made a production plan with solid buffers after each milestone, just in case we would slip.'' - \textbf{11 Bit Studios} ``Anomaly Warzone Earth'' Postmortem}
\end{quotation}

\input{threats}

%% file: threats.tex
\subsection{Limitations}
\label{sec:threats}

There are some limitations in this work. First, it's hard to generalize with a sample of 20 postmortems but, nevertheless, it shows good clues and a starting point. Second, all the postmortems are gathered only from Gamasutra portal where all related games were successful delivered. Finally, as pointed by Washburn \textit{et al.} \cite{Washburn:2016}, some postmortems' authors may be hided the true information.

%% file: relatedwork.tex
\section{Related work}
\label{sec:relatedwork}

Several researchers studied game development processes in the past years. O'Hagan \textit{et al.} recently published two studies. First, they produced a Systematic Literature Review (SLR) of the software processes used in game development \cite{OsborneOHagan2014}. They analyzed  404 papers (73\% of studies were non-indutrial), extracting 23 process models. They found that 47\% of these models are agile and 53\% are hybrid processes. In addition, they concluded that there is not one best model for game development. In their next work, they \cite{OHagan2015} investigated software process impacts on the game development, in a case study using Scrum methodology. More than conclusions, this work discusses several research  opportunities. In many aspects, our survey is complementary to these studies because we analyzed 20 industrial projects and used a different source (postmortems) to get our research data.

In another recent study, Murphy-Hill \textit{et al.} \cite{Murphy-Hill2014} interviewed 14 developers and collected 364 survey responses about game development activities. There results suggest that games have significant differences from traditional software development. They defend that game development process are not homogeneous, but instead are a rich tapestry of varying practices involving diverse people across domains.

Callele \textit{et al.} \cite{Callele2011} summarized requirements for game development and highlighted important research opportunities in this topic. Petrillo \textit{et al.} \cite{Petrillo2010} conducted a postmortem analysis study to identify good practices adopted in several game development projects, claiming that a deployment of agile methods like Scrum and XP can occur naturally, because teams already apply several agile principles in their activities. Our actual study shows that this result is valid, but agile practices are not smoothly applied as we believed at the time. Finally, Washburn \textit{et al.} \cite{Washburn:2016} analyzed 155 postmortems from 1998 till 2015 using a set of categories to highlight the best practices (``what went right'') and the pitfalls (``what went wrong'') occurred during the game development. Still, they offer recommendations to game developers.


%% file: acknowledgment.tex
\section*{Acknowledgment}

We give our thanks to the postmortems' authors for sharing their cases. This work has been partly supported by the Natural Sciences and Engineering Research Council of Canada and the Canada Research Chair on Patterns in Mixed-language Systems, and by National Council for Scientific and Technological Development of Brazil.

%% file: conclusions.tex
\section{Conclusions}
\label{sec:conclusions}

In this paper, we presented a survey of software engineering processes in video game industry from postmortem project analyses. We analyzed 20 postmortems from Gamasutra Portal and extracted their processes. As a result, we found that \textbf{iterative practices} are increasing their adoption and are applied in at least \textbf{55\% of projects}. Moreover, \textbf{agile practices} are explicitly adopted in \textbf{45\% of projects}. \textbf{Waterfall} process is still applied in \textbf{30\% of the projects}. 

These results have several conclusions and implications. First, \textbf{the ``old days" are gone, but not completely}. Despite some particularities, \textbf{this work shows that video game  and traditional software development share similar processes and practices}. The video game industry has improved its processes, adopting software engineering techniques. In the same direction of previous academic studies, \textbf{we found that iterative processes are actually \textit{mainstream} in video game industry and agile practices adoption is increasing in the last years}. However, we believe that iterative process and agile practice benefits are yet \textbf{misunderstood} by some game developers, managers, producers, publishers, and educators. To address this issue, \textbf{we suggest that educators include game agile methodologies in their regular game development courses}.

In future work, we plan two studies. First, we plan to evaluate if there is a correlation between actual project challenges and the processes and practices discussed in this work. Secondly, we plan to survey the postmortem authors, discussing about their projects to get more details and to evaluate our models.

%% file: Game.Process.Survey.GAS2016.bbl
\begin{thebibliography}{10}

\bibitem{Al-azawi2014}
R.~Al-azawi, A.~Ayesh, and M.~A. Obaidy.
\newblock {Towards Agent-based Agile approach for Game Development
  Methodology}.
\newblock In {\em 2014 World Congress on Computer Applications and Information
  Systems (WCCAIS)}, pages 1--6. IEEE, jan 2014.

\bibitem{beck2000extreme}
K.~Beck.
\newblock {\em Extreme programming explained: embrace change}.
\newblock Addison-Wesley Professional, 2000.

\bibitem{postmortem}
A.~Birk, T.~Dingsoyr, and T.~Stalhane.
\newblock Postmortem: never leave a project without it.
\newblock {\em IEEE Software}, 19, may 2002.

\bibitem{glossary6}
S.~Breakfast.
\newblock Scrum glossary: 62 scrum related terms in 50 words or less (each),
  2016.
\newblock [Online; accessed 19-January-2016].

\bibitem{callele}
D.~Callele, E.~Neufeld, and K.~Schneider.
\newblock Requirements engineering and the creative process in the video game
  industry.
\newblock In {\em 13th IEEE International Conference on Requirements
  Engineering}, August 2005.

\bibitem{Callele2011}
D.~Callele, E.~Neufeld, and K.~Schneider.
\newblock {A report on select research opportunities in requirements
  engineering for videogame development}.
\newblock In {\em 2011 Fourth International Workshop on Multimedia and
  Enjoyable Requirements Engineering (MERE'11)}, pages 26--33. IEEE, aug 2011.

\bibitem{cohn2010succeeding}
M.~Cohn.
\newblock {\em Succeeding with agile: software development using Scrum}.
\newblock Pearson Education, 2010.

\bibitem{fowler2001new}
M.~Fowler.
\newblock The new methodology.
\newblock {\em Wuhan University Journal of Natural Sciences}, 6(1-2):12--24,
  2001.

\bibitem{Fridley2013}
M.~Fridley.
\newblock {Postmortem: Kingdoms of Amalur: Reckoning}, 2013.

\bibitem{game_plan}
A.~Gershenfeld, M.~Loparco, and C.~Barajas.
\newblock {\em Game plan: the insider's guide to breaking in and succeeding in
  the computer and video game business}.
\newblock St. Martin' s Griffin Press, New York, 2003.

\bibitem{goodbye_postmortem}
W.~Hamann.
\newblock Goodbye postmortems, hello critical stage analysis.
\newblock {\em Gamasutra - The Art \& Business of Making Games}, July 2003.

\bibitem{swebok2014}
Ieee.
\newblock {\em {Guide to the software engineering body of knowledge version
  3.0}}.
\newblock 2014.

\bibitem{glossary1}
Innolution.
\newblock Agile glossary definitions, 2016.
\newblock [Online; accessed 19-January-2016].

\bibitem{Washburn:2016}
M.~W. Jr, P.~Sathiyanarayanan, M.~Nagappan, T.~Zimmermann, and C.~Bird.
\newblock ``what went right and what went wrong'': An analysis of 155
  postmortems from game development.
\newblock ACM – Association for Computing Machinery, May 2016.

\bibitem{Kanode2009}
C.~M. Kanode and H.~M. Haddad.
\newblock {Software Engineering Challenges in Game Development}.
\newblock In {\em 2009 Sixth International Conference on Information
  Technology: New Generations}, pages 260--265. IEEE, 2009.

\bibitem{Keith2010}
C.~Keith.
\newblock {\em {Agile Game Development with Scrum}}.
\newblock Addison-Wesley Professional, 1st edition, 2010.

\bibitem{Kruchten:2000:RUP}
P.~Kruchten.
\newblock {\em The Rational Unified Process: An Introduction, Second Edition}.
\newblock Addison-Wesley Longman Publishing Co., Inc., Boston, MA, USA, 2nd
  edition, 2000.

\bibitem{kruchten2004rational}
P.~Kruchten.
\newblock {\em The rational unified process: an introduction}.
\newblock Addison-Wesley Professional, 2004.

\bibitem{glossary2}
A.~L. Labs.
\newblock The agile dictionary, 2016.
\newblock [Online; accessed 19-January-2016].

\bibitem{larman2003iterative}
C.~Larman and V.~R. Basili.
\newblock Iterative and incremental development: A brief history.
\newblock {\em Computer}, (6):47--56, 2003.

\bibitem{martin2003agile}
R.~C. Martin.
\newblock {\em Agile software development: principles, patterns, and
  practices}.
\newblock Prentice Hall PTR, 2003.

\bibitem{martin2009clean}
R.~C. Martin.
\newblock {\em Clean code: a handbook of agile software craftsmanship}.
\newblock Pearson Education, 2009.

\bibitem{Murphy-Hill2014}
E.~Murphy-Hill, T.~Zimmermann, and N.~Nagappan.
\newblock {Cowboys, ankle sprains, and keepers of quality: how is video game
  development different from software development?}
\newblock In {\em Proceedings of the 36th International Conference on Software
  Engineering - ICSE 2014}, pages 1--11, New York, New York, USA, 2014. ACM
  Press.

\bibitem{pma}
M.~Myllyaho, O.~Salo, J.~Kääriäinen, J.~Hyysalo, and J.~Koskela.
\newblock A review of small and large post-mortem analysis methods.
\newblock In {\em IEEE France}, Paris, November 2004. 17th International
  Conference Software \& Systems Engineering and their Applications.

\bibitem{OHagan2015}
A.~O. O'Hagan and R.~V. O'Connor.
\newblock {Towards an Understanding of Game Software Development Processes: A
  Case Study}.
\newblock In D.~Winkler, R.~V. O’Connor, and R.~Messnarz, editors, {\em
  Communications in Computer and Information Science}, volume 301 of {\em
  Communications in Computer and Information Science}, pages 3--16. Springer
  Berlin Heidelberg, Berlin, Heidelberg, 2015.

\bibitem{Omg2011}
O.~M.~G. Omg, R.~Parida, and S.~Mahapatra.
\newblock {Business Process Model and Notation (BPMN) Version 2.0}.
\newblock Technical Report January, 2011.

\bibitem{OsborneOHagan2014}
A.~O. O’Hagan, G.~Coleman, and R.~V. O’Connor.
\newblock Software development processes for games: a systematic literature
  review.
\newblock {\em Systems, Software and Services Process Improvement}, pages
  182--193, 2014.

\bibitem{Petrillo2010}
F.~Petrillo and M.~Pimenta.
\newblock {Is agility out there? Agile Practices in Game Development}.
\newblock In {\em Proceedings of the 28th ACM International Conference on
  Design of Communication - SIGDOC '10}, page~9, New York, New York, USA, 2010.
  ACM Press.

\bibitem{Petrillo2009}
F.~Petrillo, M.~Pimenta, F.~Trindade, and C.~Dietrich.
\newblock {What went wrong? A survey of problems in game development}.
\newblock {\em Computers in Entertainment}, 7(1):1, feb 2009.

\bibitem{royce1970managing}
W.~W. Royce.
\newblock Managing the development of large software systems.
\newblock In {\em proceedings of IEEE WESCON}, volume~26, pages 328--388. Los
  Angeles, 1970.

\bibitem{glossary5}
Scrum.org.
\newblock Professional scrum developer glossary, 2016.
\newblock [Online; accessed 19-January-2016].

\bibitem{ieeeglossary1990}
A.~September.
\newblock Ieee standard glossary of software engineering terminology.
\newblock {\em Office}, 121990(1):1, 1990.

\bibitem{shore2007art}
J.~Shore et~al.
\newblock {\em The art of agile development}.
\newblock " O'Reilly Media, Inc.", 2007.

\bibitem{glossary3}
SolutionsIQ.
\newblock Agile glossary, 2016.
\newblock [Online; accessed 19-January-2016].

\bibitem{glossary4}
Telerik.
\newblock Agile terms glossary, 2016.
\newblock [Online; accessed 19-January-2016].

\end{thebibliography}
